\documentclass[12pt]{article}
\usepackage{epic}
\usepackage{epsfig}

\begin{document}

\addtolength{\baselineskip}{0.5\baselineskip}

\title{\textbf{Universal Factorization of $3n-j\ (j > 2)$ Symbols of the First and Second
Kinds for $SU(2)$ Group and Their Direct and Exact Calculation and
Tabulation}}
\author{Liqiang Wei and Alexander Dalgarno\\
Institute for Theoretical Atomic, Molecular and Optical Physics\\
Harvard University, Cambridge, MA 02138}

\maketitle

\begin{abstract}
\vspace{0.05in}

We show that general $3n-j\ (n>2)$ symbols of the first kind and
the second kind for the group $SU(2)$ can be reformulated in terms
of binomial coefficients. The proof is based on the graphical
technique established by Yutsis, et al. and through a definition
of a reduced $6-j$ symbol. The resulting $3n-j$ symbols thereby
take a combinatorial form which is simply the product of two
factors. The one is an integer or polynomial which is the single
sum over the products of reduced $6-j$ symbols. They are in the
form of summing over the products of binomial coefficients. The
other is a multiplication of all the triangle relations appearing
in the symbols, which can also be rewritten using binomial
coefficients. The new formulation indicates that the intrinsic
structure for the general recoupling coefficients is much nicer
and simpler, which might serves as a bridge for the study with
other fields. Along with our newly developed algorithms, this also
provides a basis for a direct, exact and efficient calculation or
tabulation of all the $3n-j$ symbols of the $SU(2)$ group for all
range of quantum angular momentum arguments. As an illustration,
we present the results for the $12-j$ symbols of the first kind.
\end{abstract}

\vspace{0.35in}

\section{Introduction}

The quantum theory of angular momentum is a fundamental field in
sciences. In particular, the topic of angular momentum coupling
scheme, which is an indication of the geometric aspect of an
interacting many-body system, plays a paramount role in a variety
of disciplines [1-4]. In the mathematical front line of
development, there exists a close relation for the coupling or
recoupling coefficients of angular momenta with its various
branches such as combinatorial analysis, special functions,
calculus of finite difference, algebra, and topology [4]. The
investigation of their properties and relations should always have
great implications or impacts to those fields. For example, the
Clebsch-Gordan coefficients, Racah coefficients and other
coefficients are expressible in terms of generalized
hypergeometric functions of many variables [1,4,5]. These
functions are related to the discrete orthogonal polynomial, which
have wide applications in the areas from numerical analysis, to
solutions of differential equation in physics, and to topological
study. The applications of the vector coupling coefficients also
abound in physical sciences including the fields of quantum
chemistry and quantum molecular dynamics. The Clebsch-Gordan
coefficients and Racah coefficients, for instance, are the basic
mathematical apparatus in the formulation of rovibrational
spectra, quantum scattering process, and photodissociation
dynamics for polyatomic molecules or molecular complexes [3,7-9].
They are also very useful for the evaluation of molecular
integrals in electronic structure calculation, where the primary
basis functions are chosen to be the spherical harmonics or solid
harmonics [10]. The myriad applications of coupling theory of
quantum angular momentum in the traditional fields of atomic
physics, nuclear physics, and elementary particle physics are
well-known. The noticeable recent development is their role in the
study of quantum integrable systems. One example is that the
products of two wavefunctions for a Calogero-Sutherland system
with a potential $v(x)=sin^{-2}x$ is proved to be identical with
the Clebsch-Gordan coefficients [11].

In most of the applications described above, the exact
determination of angular momentum coupling and recoupling
coefficients for all ranges of quantum angular momentum arguments
is often very critical. In the first place, it is obvious that the
domain of definition for these quantum numbers occurring in the
coefficients should be full range for the investigation in
mathematical applications. Semiclassical limits of coupling and
recoupling coefficients also play very important roles in modern
physics. This includes famous Ponzano and Regge's relation of the
asymptotic $6-j$ symbols with the partition function for three
dimensional quantum gravity [12]. The Heisenberg correspondence
limits or asymptotic approximations of these coefficients are also
useful in the study of scattering cross section and Rydberg states
of molecules [13,14]. Furthermore, in quantum molecular scattering
study for the chemical reactions, the angular momentum coupling
coefficients naturally arise in the evaluation of interaction
matrix elements between the channels corresponding to different
chemical arrangements. When the collision energy increases, the
channels with large angular momentum become energetically
accessible, and the accurate computation of their coupling
coefficients is important for a correct description of the
chemical dynamic processes [15].

Recently, we have independently shown that all the $3-j$, $6-j$,
and $9-j$ symbols can be reformulated as a common combinatorial
form by utilization of binomial coefficients [16]. The intrinsic
structure of these symbols is found to be much nicer and simpler
than thought before, which is simply the product of a polynomial
and a square root of integer multiplication and division. This
opens an avenue for the symbolic study of these symbols, and also
lays the foundation for an exact, direct and fast numerical
calculation or tabulation of each coupling or recoupling
coefficient. In addition, it provides a convenient numerical
approach for exactly locating all the structural and non-trivial
zeros of angular momentum coupling or recoupling coefficients
since these zeros can be determined exclusively by the polynomial
part from our schemes. Although it is still an interesting work to
identify their physical implication, few non-trivial zeros do show
obvious associated physical meaning [1].

This paper aims at a generalization of methods we developed
previously for the work on the direct and exact calculation and
tabulation of $3-j$, $6-j$, and $9-j$ symbols to the case of
$3n-j\ (n>2)$ symbols of the first kind and the second kind [16].
The concentration is more on the analysis of their intrinsic
structure. In Section 2, we begin with a review of the general
theory about $3n-j$ symbols, their calculation, and their
decomposition in particular. By utilizing the graphic technique,
developed by Yutsis, et al., and via a definition of reduced $6-j$
symbols, we give a formal description and detailed proof about the
statement that $\it{all\ the\ 3n-j\ (n>2)\ symbols\ of\ the\
first}$\\ $\it{kind\ and\ the\ second\ kind\ can\ be\ factorized\
into\ a\ prefactor\ and\ an\ integer}$\\ $\it{or\ a\ polynomial\
part}$. The latter is simply the single sum over the products of
reduced $6-j$ symbols. In Section 3, we concretize our development
in Section 2 and give the calculation results for the $12-j$
symbols of the first kind. The Section 4 is a discussion and
conclusion.

\vspace{0.35in}

\section{Factorization of $3n-j$ Symbols with Binomial
Coefficients}

The angular momentum coupling leading to a total angular momentum
of a composite system can be characterized by a binary tree. Each
binary coupling scheme corresponds to a way for constructing the
basis vectors for the tensor product Hilbert space. The general
angular momentum recoupling coefficients $3n-j$ for a system with
$n+1$ degrees of freedom are the unitary transformation matrix
elements connecting these different pairwise and sequential
angular momentum coupling schemes. These coefficients have been
formulated in different manners including explicit algebraic
expressions, the formulae in terms of Clebsch-Gordan coefficients
or Racah coefficients, and graphic representations, etc., serving
for different purposes [2]. From the computational point of view,
however, the forms of these different expressions really matter
considering the requirements of accuracy, efficiency, or
convenience for their practical evaluation. For example, the
explicit algebraic expressions were most often used in the
numerical calculation in the early years. But it is well-known
that they suffer serious numerical instability and overflow or
underflow are occurring issues. The recursive relations for the
coupling or recoupling coefficients are used in the accurate
calculation. However, it is not a direct and efficient approach
either [17,18]. Other methods based on the hyper-geometric series
expansions or graphical techniques have been developed, but the
programs are usually too complicated to be practical. The
representation of the general recoupling coefficients in terms of
$6-j$ symbols are also being employed for the computation, but
similar problems exists as those using algebraic expressions [4].
Recently, we have developed an approach for direct and exact
computation and tabulation of $3-j$, $6-j$, and $9-j$ symbols for
all ranges of angular momentum arguments. The calculation scheme
is based on the reformulation of these symbols by utilization of
binomial coefficients. The resulting formulae are simply the
product of two factors, which lay the foundation for a symbolic
study as well as for a direct and exact calculation of its
numerical values. Furthermore, it gives a deeper understanding of
the intrinsic structure of these coupling or recoupling
coefficients, and establishes a link for the study with the other
branches of the fields. Encouraged by this work, we believe that
the statement that a recoupling coefficient can be factorized into
an integer or a polynomial part times a multiplication of all the
triangle relations appearing in the coefficient might generally
holds for any number or any kind of angular momentum coupling.
Motivated by getting a better understanding of their intrinsic
structure including their connection to other fields, and giving a
unified exact calculation scheme, it is very natural thing to
extend previous work to the case of the general $3n-j$ symbols.

In our previous work for the $3-j$ and $6-j$ symbols, the
derivation is from their algebraic expressions. For $9-j$ symbols,
however, the derivation is from their expression in terms of $6-j$
symbols rather than the explicit triple-sum formula. In fact, for
the general $3n-j\ (n>2)$ recoupling coefficients, which includes
$9-j$ symbols as a special case, there is a well-known fundamental
theorem which states that they can be written as multiple sums
over the products of $6-j$ symbols. The soundness of this theorem
is based on the recognition that there are only two basic
operations: commutation or association in the binary coupling
theory which can transfer one binary coupling scheme to another.
The operation of sequences of these two operations will either
change the phase of basis vectors or introduce additional factor
of Racah coefficients or $6-j$ symbols. From the derivation for
$9-j$ symbols, it seems obvious that a general $3n-j$ symbol can
be reformulated in terms of reduced $6-j$ symbols defined in [16]
similar to that for $9-j$ symbols. However, an unpleasant fact
about these transformation is that there exist different paths
from a given binary coupling scheme to another. In addition, what
kind of actual form to which a recoupling coefficient can be
reduced depends on the type of that coefficient. For the special
types, the $3n-j$ symbols of the first kind and the second kind,
which are most commonly used and most important up to present
time, things look much obvious and we do achieve our goals here.

The $3n-j$ symbols of the first kind are proportional to the
unitary transformation matrix elements between two basis vectors
associated with two different coupling schemes, defined by
\begin{eqnarray}
\nonumber \left\{\begin{array}{ccccccc}j_{1}&&j_{2}&&...&j_{n}&\\
&k_{1}&&k_{2}&...&&k_{n}\\l_{1}&&l_{2}&&...&l_{n}&\end{array}\right\}_{1}\equiv
 \frac{(-1)^{j_{1}+l_{n}-j_{n}-l_{1}}}{\left[\Pi_{\alpha=2}^{n}(2j_{\alpha}+1)
 (2l_{\alpha}+1)\right]^{1/2}}\times \\
\times \left<
\begin{picture}(80,45)
\put(5,30){\circle{4}} \put(15,20){\circle*{4}}
\put(17,18){\circle*{1}}\put(19,16){\circle*{1}}\put(21,14){\circle*{1}}\put(23,12){\circle*{1}}
\put(25,10){\circle*{4}}\put(35,0){\circle*{4}}
\put(37,-2){\circle*{1}}\put(39,-4){\circle*{1}}\put(41,-6){\circle*{1}}\put(43,-8){\circle*{1}}
\put(45,-10){\circle*{4}} \put(55,-20){\circle*{4}}
\put(5,30){\circle{4}}\put(25,30){\circle{4}}
\put(35,20){\circle{4}}
\put(45,10){\circle{4}}\put(55,0){\circle{4}}\put(65,-10){\circle{4}}
\put(5,30){\line(1,-1){10}}\put(25,10){\line(1,-1){10}}\put(45,-10){\line(1,-1){10}}
\put(15,20){\line(1,1){10}}\put(25,10){\line(1,1){10}}\put(35,0){\line(1,1){10}}
\put(45,-10){\line(1,1){10}}\put(55,-20){\line(1,1){10}}
\put(5,38){$j_{1}$}\put(10,10){$j_{2}$}\put(20,0){$j_{i}$}\put(30,-10){$j_{i+1}$}
\put(40,-20){$j_{n}$}\put(50,-33){$k_{n}$}
\put(30,30){$k_{1}$}\put(40,20){$k_{i}$}
\put(50,10){$k_{i+1}$}\put(60,0){$k_{n-1}$} \put(70,-10){$l_{1}$}
\end{picture}
\begin{picture}(2,45) \put(1,-20){\line(0,1){50}} \end{picture}
\begin{picture}(80,45)\put(5,30){\circle{4}} \put(15,20){\circle*{4}}
\put(17,18){\circle*{1}}\put(19,16){\circle*{1}}\put(21,14){\circle*{1}}\put(23,12){\circle*{1}}
\put(25,10){\circle*{4}}\put(35,0){\circle*{4}}
\put(37,-2){\circle*{1}}\put(39,-4){\circle*{1}}\put(41,-6){\circle*{1}}\put(43,-8){\circle*{1}}
\put(45,-10){\circle*{4}} \put(55,-20){\circle*{4}}
\put(5,30){\circle{4}}\put(25,30){\circle{4}}
\put(35,20){\circle{4}}
\put(45,10){\circle{4}}\put(55,0){\circle{4}}\put(65,-10){\circle{4}}
\put(5,30){\line(1,-1){10}}\put(25,10){\line(1,-1){10}}\put(45,-10){\line(1,-1){10}}
\put(15,20){\line(1,1){10}}\put(25,10){\line(1,1){10}}\put(35,0){\line(1,1){10}}
\put(45,-10){\line(1,1){10}}\put(55,-20){\line(1,1){10}}
\put(5,38){$l_{1}$}\put(10,10){$l_{2}$}\put(20,0){$l_{i+1}$}\put(30,-10){$l_{i+2}$}
\put(40,-20){$l_{n}$}\put(50,-33){$k_{n}$}
\put(30,30){$k_{1}$}\put(40,20){$k_{i}$}
\put(50,10){$k_{i+1}$}\put(60,0){$k_{n-1}$} \put(70,-10){$j_{1}$}
\end{picture}
\right>,
\end{eqnarray}
\\
\\
where the phase and proportional coefficients are introduced so
that the symbols possess the maximal symmetry. They contain the
$6-j$ symbols and the $9-j$ symbols as special cases except for
possible phase factors. Similarly, the $3n-j$ symbols of the
second kind are related to the transformation matrix elements
between two different coupling schemes in the following way,
\begin{eqnarray}
\nonumber \left\{\begin{array}{ccccccc}j_{1}&&j_{2}&&...&j_{n}&\\
&k_{1}&&k_{2}&...&&k_{n}\\l_{1}&&l_{2}&&...&l_{n}&\end{array}\right\}_{2}\equiv
 \frac{(-1)^{j_{2}-j_{1}-k_{1}+k_{n}-l_{n}-l_{1}}}{\left[(k_{1}+1)(k_{n}+1)
 \Pi_{\alpha=2}^{n-1}(2j_{\alpha}+1)
 (2l_{\alpha+1}+1)\right]^{1/2}}\times && \\
 \times\left<
\begin{picture}(80,50)\put(5,30){\circle{4}} \put(15,20){\circle*{4}}
\put(17,18){\circle*{1}}\put(19,16){\circle*{1}}\put(21,14){\circle*{1}}\put(23,12){\circle*{1}}
\put(25,10){\circle*{4}}\put(35,0){\circle*{4}}
\put(37,-2){\circle*{1}}\put(39,-4){\circle*{1}}\put(41,-6){\circle*{1}}\put(43,-8){\circle*{1}}
\put(45,-10){\circle*{4}} \put(55,-20){\circle*{4}}
\put(5,30){\circle{4}}\put(25,30){\circle{4}}
\put(35,20){\circle{4}}
\put(45,10){\circle{4}}\put(55,0){\circle{4}}\put(65,-10){\circle{4}}
\put(5,30){\line(1,-1){10}}\put(25,10){\line(1,-1){10}}\put(45,-10){\line(1,-1){10}}
\put(15,20){\line(1,1){10}}\put(25,10){\line(1,1){10}}\put(35,0){\line(1,1){10}}
\put(45,-10){\line(1,1){10}}\put(55,-20){\line(1,1){10}}
\put(5,38){$l_{2}$}\put(10,10){$k_{1}$}\put(20,0){$j_{i}$}\put(30,-10){$j_{i+1}$}
\put(40,-20){$j_{n-1}$}\put(50,-33){$j_{n}$}
\put(30,30){$l_{1}$}\put(40,20){$k_{i-1}$}
\put(50,10){$k_{i}$}\put(60,0){$k_{n-2}$} \put(70,-10){$k_{n-1}$}
\end{picture}
\begin{picture}(2,50) \put(1,-20){\line(0,1){50}} \end{picture}
\begin{picture}(80,50)\put(5,30){\circle{4}} \put(15,20){\circle*{4}}
\put(17,18){\circle*{1}}\put(19,16){\circle*{1}}\put(21,14){\circle*{1}}\put(23,12){\circle*{1}}
\put(25,10){\circle*{4}}\put(35,0){\circle*{4}}
\put(37,-2){\circle*{1}}\put(39,-4){\circle*{1}}\put(41,-6){\circle*{1}}\put(43,-8){\circle*{1}}
\put(45,-10){\circle*{4}} \put(55,-20){\circle*{4}}
\put(5,30){\circle{4}}\put(25,30){\circle{4}}
\put(35,20){\circle{4}}
\put(45,10){\circle{4}}\put(55,0){\circle{4}}\put(65,-10){\circle{4}}
\put(5,30){\line(1,-1){10}}\put(25,10){\line(1,-1){10}}\put(45,-10){\line(1,-1){10}}
\put(15,20){\line(1,1){10}}\put(25,10){\line(1,1){10}}\put(35,0){\line(1,1){10}}
\put(45,-10){\line(1,1){10}}\put(55,-20){\line(1,1){10}}
\put(5,38){$l_{2}$}\put(10,10){$l_{3}$}\put(20,0){$l_{i}$}\put(30,-10){$l_{i+1}$}
\put(40,-20){$k_{n}$}\put(50,-33){$j_{n}$}
\put(30,30){$k_{2}$}\put(40,20){$k_{i-1}$}
\put(50,10){$k_{i}$}\put(60,0){$l_{1}$} \put(70,-10){$j_{1}$}
\end{picture}
 \right>.
\end{eqnarray}
\\
\\
The study of their properties such as reduction is best performed
through the graphical technique developed by Yutsis, Levinson, and
Vanagas [19-23]. It merits a simple and clear presentation of
angular momentum relations and facilitates their general analysis.
In the graphical method, a general $3n-j$ symbol is represented by
a closed diagram or polygon with $3n$ line segments representing
the angular momenta and the $2n$ vertices being their
Clebsch-Gordan coupling coefficients. It is formed by combining
the free ends of the line segments with the same angular momentum
arguments for the coupling coefficients. For the reduction process
we are studying, however, it is an opposite process, which
corresponds to the separation of a diagram into its several
sub-diagrams. Some rules have been developed guiding this
decomposition, which can be classified into two types of
situation. The first one is the case when the diagram is separable
on one, two, or three lines. The diagram can therefore be
decomposed into two independent parts. The second one is the
situation when the diagram is separable on four or more lines. The
decomposition of this diagram will bring one or more summation
indices, depending on the number of connected lines. The reduction
process of an angular momentum diagram is a repeating and
sequential application of above rules. This also leads to the
fundamental theorem in quantum angular momentum theory we
mentioned before.

For the $3n-j$ symbols of the first kind we are studying, their
diagarm is a Mobius band or a band $\it{with}$ braiding, while the
diagram for the second kind is the band $\it{without}$ braiding as
shown in reference [24]. The common feature of these diagrams is
that after utilization of orthogonal relation for the coupled
states, which introduces a summation index, they both become the
ones separable on three lines. Repeating and sequential
application of the rule for the diagram separable on three lines
then leads to the complete reduction of the diagram, for example,
for the $3n-j$ symbols of the first kind, in terms of the diagrams
for the $6-j$ symbols. Analytically, this process indicates the
following relation,
\begin{eqnarray}
\nonumber \left\{\begin{array}{ccccccc}j_{1}&&j_{2}&&...&j_{n}&\\
&k_{1}&&k_{2}&...&&k_{n}\\l_{1}&&l_{2}&&...&l_{n}&\end{array}\right\}_{1}=
\sum_{x}(-1)^{S+(n-1)x}(2x+1)\left\{\begin{array}{ccc}j_{1}&l_{1}&x\\
l_{2}&j_{2}&k_{1}\end{array}\right\}  \times\\
\times \left\{\begin{array}{ccc}j_{2}&l_{2}&x\\
l_{3}&j_{3}&k_{2}\end{array}\right\}...
\left\{\begin{array}{ccc}j_{n-1}&l_{n-1}&x\\
l_{n}&j_{n}&k_{n-1}\end{array}\right\}
\left\{\begin{array}{ccc}j_{n}&l_{n}&x\\
j_{1}&l_{1}&k_{n}\end{array}\right\} ,
\end{eqnarray}
which is just the single sum over the products of $6-j$ symbols.
The same procedure also yields the relation for the $3n-j$ symbols
of the second kind,
\begin{eqnarray}
\nonumber \left\{\begin{array}{ccccccc}j_{1}&&j_{2}&&...&j_{n}&\\
&k_{1}&&k_{2}&...&&k_{n}\\l_{1}&&l_{2}&&...&l_{n}&\end{array}\right\}_{2}=
\sum_{x}(-1)^{S+nx}(2x+1)\left\{\begin{array}{ccc}j_{1}&l_{1}&x\\
l_{2}&j_{2}&k_{1}\end{array}\right\}  \times\\
\times \left\{\begin{array}{ccc}j_{2}&l_{2}&x\\
l_{3}&j_{3}&k_{2}\end{array}\right\}...
\left\{\begin{array}{ccc}j_{n-1}&l_{n-1}&x\\
l_{n}&j_{n}&k_{n-1}\end{array}\right\}
\left\{\begin{array}{ccc}j_{n}&l_{n}&x\\
l_{1}&j_{1}&k_{n}\end{array}\right\} ,
\end{eqnarray}
where
\begin{equation}
  S = \sum_{i=1}^{n} (j_{i}+k_{i}+l_{i}).
\end{equation}
Two points are noticeable in the above diagram decomposition.
First, any triangle relation arising from the diagram separation
will always appear in both separated parts. Second, for the $3n-j$
symbols of the first and second kinds, all the $6-j$ symbols are
symmetric in the sense that summation is single fold and only one
summation index appears in each of the $6-j$ symbols. These are
critical in current reformulation.

Now, we define an integer factor or polynomial by
\begin{eqnarray}
\nonumber\left[\begin{array}{ccc}j_{1}&j_{2}&j_{3}\\j_{4}&j_{5}&j_{6}\end{array}\right]\equiv
\sum_{k}(-1)^{k}\left(\begin{array}{c}k+1\\k-j_{1}-j_{2}-j_{3}\end{array}\right)
\left(\begin{array}{c}j_{1}+j_{2}-j_{3}\\k-j_{3}-j_{4}-j_{5}\end{array}\right)\times\\
\times\left(\begin{array}{c}j_{1}-j_{2}+j_{3}\\k-j_{2}-j_{4}-j_{6}\end{array}\right)
\left(\begin{array}{c}-j_{1}+j_{2}+j_{3}\\k-j_{1}-j_{5}-j_{6}\end{array}\right),
\end{eqnarray}
which is the sum over the products of four binomial coefficients.
It relates to the $6-j$ symbol in the following way,
\begin{equation}
\left\{\begin{array}{ccc}j_{1}&j_{2}&j_{3}\\j_{4}&j_{5}&j_{6}\end{array}\right\}=
\frac{\Delta(j_{3}j_{4}j_{5})\Delta(j_{2}j_{4}j_{6})\Delta(j_{1}j_{5}j_{6})}{\Delta(j_{1}j_{2}j_{3})}
\left[\begin{array}{ccc}j_{1}&j_{2}&j_{3}\\j_{4}&j_{5}&j_{6}\end{array}\right].
\end{equation}
It is simply the polynomial part of the $6-j$ symbol but
nevertheless has its full symmetry. In addition, the triangle
relation appearing in the denominator can be any one of the four
triangle relations for the $6-j$ symbol. Because of its status as
a fundamental building block in all following formulation, we call
it the $\it{reduced}$ $\it{6-j}$ $\it{symbol}$. Substitution of
Eq. (7) into Eq. (3) or (4) yields the following two formulae,
\begin{eqnarray}
\nonumber \left\{\begin{array}{ccccccc}j_{1}&&j_{2}&&...&j_{n}&\\
&k_{1}&&k_{2}&...&&k_{n}\\l_{1}&&l_{2}&&...&l_{n}&\end{array}\right\}_{1}=
\Delta(j_{1}j_{2}k_{1})\Delta(l_{2}l_{1}k_{1})\Delta(j_{2}j_{3}k_{2})\Delta(l_{3}l_{2}k_{2})...\times \\
\times\Delta(j_{n-1}j_{n}k_{n-1})\Delta(l_{n}l_{n-1}k_{n-1})\Delta(j_{n}l_{1}k_{n})
\Delta(j_{1}l_{n}k_{n})\left[\begin{array}{ccccccc}j_{1}&&j_{2}&&...&j_{n}&\\
&k_{1}&&k_{2}&...&&k_{n}\\l_{1}&&l_{2}&&...&l_{n}&\end{array}\right]_{1},
\end{eqnarray}
where
\begin{eqnarray}
\nonumber \left[\begin{array}{ccccccc}j_{1}&&j_{2}&&...&j_{n}&\\
&k_{1}&&k_{2}&...&&k_{n}\\l_{1}&&l_{2}&&...&l_{n}&\end{array}\right]_{1}\equiv
\sum_{x}(-1)^{S+(n-1)x}(2x+1)\left[\begin{array}{ccc}j_{1}&l_{1}&x\\
l_{2}&j_{2}&k_{1}\end{array}\right]  \times\\
\times \left[\begin{array}{ccc}j_{2}&l_{2}&x\\
l_{3}&j_{3}&k_{2}\end{array}\right]...
\left[\begin{array}{ccc}j_{n-1}&l_{n-1}&x\\
l_{n}&j_{n}&k_{n-1}\end{array}\right]
\left[\begin{array}{ccc}j_{n}&l_{n}&x\\
j_{1}&l_{1}&k_{n}\end{array}\right] ,
\end{eqnarray}
and
\begin{eqnarray}
\nonumber \left\{\begin{array}{ccccccc}j_{1}&&j_{2}&&...&j_{n}&\\
&k_{1}&&k_{2}&...&&k_{n}\\l_{1}&&l_{2}&&...&l_{n}&\end{array}\right\}_{2}=
\Delta(j_{1}j_{2}k_{1})\Delta(l_{2}l_{1}k_{1})\Delta(j_{2}j_{3}k_{2})\Delta(l_{3}l_{2}k_{2})...\times \\
\times\Delta(j_{n-1}j_{n}k_{n-1})\Delta(l_{n}l_{n-1}k_{n-1})\Delta(j_{n}l_{1}k_{n})
\Delta(j_{1}l_{n}k_{n})\left[\begin{array}{ccccccc}j_{1}&&j_{2}&&...&j_{n}&\\
&k_{1}&&k_{2}&...&&k_{n}\\l_{1}&&l_{2}&&...&l_{n}&\end{array}\right]_{2},
\end{eqnarray}
where
\begin{eqnarray}
\nonumber \left[\begin{array}{ccccccc}j_{1}&&j_{2}&&...&j_{n}&\\
&k_{1}&&k_{2}&...&&k_{n}\\l_{1}&&l_{2}&&...&l_{n}&\end{array}\right]_{2}\equiv
\sum_{x}(-1)^{S+nx}(2x+1)\left[\begin{array}{ccc}j_{1}&l_{1}&x\\
l_{2}&j_{2}&k_{1}\end{array}\right]  \times\\
\times \left[\begin{array}{ccc}j_{2}&l_{2}&x\\
l_{3}&j_{3}&k_{2}\end{array}\right]...
\left[\begin{array}{ccc}j_{n-1}&l_{n-1}&x\\
l_{n}&j_{n}&k_{n-1}\end{array}\right]
\left[\begin{array}{ccc}j_{n}&l_{n}&x\\
l_{1}&j_{1}&k_{n}\end{array}\right].
\end{eqnarray}

   Eqs. (8) to (11) constitute the final results of this Section.
By utilization of binomial coefficients, the $3n-j\ (n>2)$ symbols
of the first and second kinds are now in a separated form with one
factor being the product of all the triangle relations occurring
in the symbol, and the other being an integer or polynomial. It is
in the form of $\it{two}$-fold summation, and the corresponding
actual form in polynomial is a topic for a further study.

\vspace{0.35in}

\section{Examples: $12-j$ Symbols of the First Kind}

As an illustration of the power of the methods and algorithms
discussed previously for the reformulation and exact calculation
of angular momentum recoupling coefficients, we present the
results for the $12-j$ symbols of the first kind. The $12-j$
symbols arises in the recoupling of five angular momenta
[2,24,25,26]. They are related to the theory of the fractional
parentage coefficients $(fpc)$, which are used, for example, in
the construction of wave functions for $N$ identical particles and
for the evaluation of their matrix elements for operators from the
counterparts for $N-1$ identical particles coupled with one more
particle [27]. Recently, it is found that the well-known Ponzano
and Regge's formula, which connects the asymptotic form of the
Racah-Wigner $6-j$ symbols to the partition function for three
dimensional quantum gravity mentioned in the beginning, can be
extended to the physically significant four-dimensional case by
utilization of the $12-j$ symbols [28]. There are five distinct
types of abstract cubic graphs associated with the coupling of
five angular momenta. However, there are three of them when the
corresponding recoupling coefficients can be directly factorized
into the products of $6-j$ symbols or $9-j$ symbols according to
the rules for the first case discussed in previous section.
Therefore, the actual number of the $12-j$ symbols is two [1,2].
The $12-j$ symbols of the first kind, for example, are the
transformation of following two coupling schemes,
\begin{eqnarray}
\nonumber \left\{\begin{array}{cccccccc}j_{1}&&j_{6}&&j_{7}&&j_{8}&\\
&j_{2}&&j_{3}&&j_{4}&&j\\j_{5}&&j^{'}_{6}&&j^{'}_{7}&&j^{'}_{8}&\end{array}\right\}_{1}\equiv
\frac{(-1)^{j_{1}-j_{5}-j_{8}+j_{8}^{'}}}{\left[(2j_{6}+1)(2j_{7}+1)(2j_{8}+1)(2j^{'}_{6}+1)
(2j^{'}_{7}+1)(2j^{'}_{8}+1)\right]^{1/2}}\times && \\
\times
<\left\{\left[\left<(j_{1}j_{2})j_{6},j_{3}\right>j_{7},j_{4}\right]j_{8},j_{5}\right\}j|
\left\{\left[\left<(j_{5}j_{2})j^{'}_{6},j_{3}\right>j^{'}_{7},j_{4}\right]j^{'}_{8},j_{1}
\right\}j>.
\end{eqnarray}
From the results of last Section, they can be formulated in the
factorized form as follows,
\begin{eqnarray}
\nonumber \left\{\begin{array}{cccccccc}j_{1}&&j_{6}&&j_{7}&&j_{8}\\
&j_{2}&&j_{3}&&j_{4}&&j\\j_{5}&&j^{'}_{6}&&j^{'}_{7}&&j^{'}_{8}&\end{array}\right\}_{1}=
\Delta(j_{1}j_{2}j_{6})\Delta(j_{6}j_{3}j_{7})\Delta(j_{7}j_{4}j_{8})\Delta(j_{8}j_{5}j)\times \\
\times\Delta(j_{5}j_{2}j^{'}_{6})\Delta(j^{'}_{6}j_{3}j^{'}_{7})\Delta(j^{'}_{7}j_{4}j^{'}_{8})
\Delta(j^{'}_{8}j_{1}j)\left[\begin{array}{cccccccc}j_{1}&&j_{6}&&j_{7}&&j_{8}\\
&j_{2}&&j_{3}&&j_{4}&&j\\j_{5}&&j^{'}_{6}&&j^{'}_{7}&&j^{'}_{8}&\end{array}\right]_{1}
\end{eqnarray}
where the integer or polynomial part is the sum over products of
reduced $6-j$ symbols defined by
\begin{eqnarray}
\nonumber \left[\begin{array}{cccccccc}j_{1}&&j_{6}&&j_{7}&&j_{8}\\
&j_{2}&&j_{3}&&j_{4}&&j\\j_{5}&&j^{'}_{6}&&j^{'}_{7}&&j^{'}_{8}&\end{array}\right]_{1}=
\sum_{x}(-1)^{S-x}(2x+1)\left[\begin{array}{ccc}j_{1}&j_{5}&x\\
j^{'}_{6}&j_{6}&j_{2}\end{array}\right]  \times\\
\times \left[\begin{array}{ccc}j_{6}&j^{'}_{6}&x\\
j^{'}_{7}&j_{7}&j_{3}\end{array}\right]
\left[\begin{array}{ccc}j_{7}&j^{'}_{7}&x\\
j^{'}_{8}&j_{8}&j_{4}\end{array}\right]
\left[\begin{array}{ccc}j_{8}&j^{'}_{8}&x\\
j_{1}&j_{5}&j\end{array}\right],
\end{eqnarray}
where the $S$, as defined before, is the sum of all twelve angular
momentum arguments. It is a function of all twelve angular
momentum arguments. However, unlike the cases for the rotation
matrices and $3-j$ and $6-j$ symbols, the actual form of the
corresponding polynomial is unknown. As already demonstrated, the
rotational matrices are expressible in terms of Jacobi polynomial
[1], the $3-j$ symbols are proportional to Hahn polynomial [4],
and the $6-j$ symbols are related to Racah polynomial [4]. The
polynomial the $9-j$ symbols correspond to with explicit
expression is still unknown also [29].

Having rewritten the $12-j$ symbols of the first kind in the
desired form, we are now in a position to calculate and tabulate
their numerical values. One of the major advantages in our
formulation is in the realization of a $\it{direct}$ and
$\it{exact}$ calculation and tabulation of $\it{all}$ recoupling
coefficients of the first and second kinds for $\it{all}$ ranges
of angular momentum arguments. In the summation step, we have
developed a series of algebraic operation routines for large
integer, where integer algebraic addition and multiplication are
performed in a basis of 32768. The array for the integer is
arranged in an order of increasing power so that the rounding-off
error can be avoided when converting to the decimal values and
final results are exact. In addition, we evaluate binomial
coefficients recursively rather than doing the calculation by
definition. This avoids the numerical errors from computation
while increases the efficiency of computation. In Table 1, we list
a table of $12-j$ symbols and corresponding decimal values.

The structural zeros and accidental zeros of the $12-j$ symbols
can also be easily and safely determined by our schemes, which are
just the zeros of the corresponding polynomial part. Obviously,
exploration of the mathematical or physical meaning of these zeros
should be an interesting and valuable investigation. We list in
Table 2 some of the structural or accidental zeros in order that
the future researchers might have a comparative study.

\vspace{0.35in}

\section{Discussion and Conclusion}

In this paper, we have factorized the $3n-j$ symbols of the first
kind and second kind in a form which is a prefactor times an
integer or a polynomial. The immediate benefit of this
reformulation is that, when combined with our developed
algorithms, it provides an analytical foundation for a direct,
exact, and fast calculation of all the $3n-j$ symbols of the first
and second kinds. It also gives a convenient and exact approach
for determining all their structural or accidental zeros. It is
really very significant considering their wide applications in a
variety of fields.

Furthermore, from the analytical points of view, the reformulation
indicates a simpler algebraic structure for the recoupling
coefficients which will serve as a bridge for the study with other
fields. The actual form of the integer or polynomial part need to
be explored. Their properties including the relations to the other
areas are expected to be investigated.

Just like the case when the building block for recoupling
coefficients has been changed from the fundamental Clebsch-Gordan
coefficients to the Racah coefficients which is indicated in the
fundamental theorem of quantum angular momentum, it is also a
surprise that the reduced $6-j$ symbol, as we have defined, can
serve as a similar building block for the $3n-j$ symbols of the
first and second kinds, where the general relation with polynomial
has been established. Nevertheless, for the general recoupling
coefficients, which cover the cases of the third or higher kinds,
things do not look so clear. The endeavor to explore the similar
structure seems elusive. Even though we can always transfer a
$3n-j$ symbol to the multiple sum over the products of $6-j$
symbols, it is by no means obvious that we might reformulate it as
the multiple sum over the products of reduced $6-j$ symbols with
one factor being an integer or polynomial. What kind of the role
the binomial coefficient plays in the current reformulation? What
is the building block for all the recoupling coefficients when
they have the desired factorization? These are the questions to be
answered. We have more discoveries to make and we will have more
to understand for their intrinsic structure.

\vspace{0.45in}
\noindent

{\bf \Large{Acknowledgment}}

\vspace{0.15in}
\noindent

This work was partially supported by the National Science
Foundation through a grant for the Institute for Theoretical
Atomic, Molecular and Optical Physics at Harvard University and
Smithsonian Astrophysical Observatory.

\newpage
\begin{center}
{\bf Table Captions}
\end{center}

\mbox{}\\
\noindent {\bf Table 1.} A table of $12-j$ symbols and
corresponding decimal values.

\mbox{}\\
\noindent {\bf Table 2.} Ten examples of structural or accidental
zeros of $12-j$ symbols with small angular momentum arguments.

\begin{table}[p]
\begin{center}
  \begin{tabular}{|l|l|l|}  \hline\hline
 $\left\{\begin{array}{cccccccc}j_{1}&&j_{6}&&j_{7}&&j_{8}&\\
&j_{2}&&j_{3}&&j_{4}&&j\\j_{5}&&j^{'}_{6}&&j^{'}_{7}&&j^{'}_{8}&\end{array}\right\}$
&\ \ \ \ \ Tabulation &\ \   Exact Decimal Values \\
\hline\hline
$\left\{ \begin{array}{llllllll} 0 && 1 && 1 && 1 & \\ & 1 && 1 && 0 && 1\\
1 && 1 && 1 && 1 & \end{array} \right\} $
 & $ \begin{array}{l} 2^{4}3^{-6} \\ \times (1);\end{array} $
 & 0.148148148148148   \\  \hline
$\left\{ \begin{array}{llllllll} 1 && 1 && 2 && 2 & \\ & 2 && 2 && 2 && 1\\
1 && 2 && 1 && 2 & \end{array} \right\} $
 & $\begin{array}{l} -2^{-2}3^{-3}5^{-8}
 7^{-1}89^{2} \\
 \times (1);
  \end{array} $ & -5.179037928267551E-003   \\  \hline
$\left\{ \begin{array}{llllllll} 2 && 1 && 2 && 1 & \\ & 2 && 2 && 1 && 2\\
1 && 1 && 2 && 2 & \end{array} \right\} $
 & $\begin{array}{l}
2^{-6}3^{1}5^{-8}7^{1}41^{2} \\
 \times (1);
  \end{array} $  & 3.757712069863789E-002  \\  \hline
$\left\{ \begin{array}{llllllll} 3 && 3 && 5 && 4 & \\ & 2 && 4 && 5 && 3\\
4 && 2 && 4 && 5 & \end{array} \right\} $
 & $\begin{array}{l} -2^{-14}3^{-16}
 5^{-7}7^{-10}\\
\times 11^{-6}13^{-3}743^{2} \\
\times (471, 15204,19117); \end{array} $ & -1.528195512735275E-003 \\
   \hline
$\left\{ \begin{array}{llllllll} 4 && 3 && 5 && 6 & \\ & 3 && 5 && 6 && 4\\
2 && 4 && 3 && 5 & \end{array} \right\} $
 & $\begin{array}{l} 3^{-14}5^{-2}
 7^{-2}11^{-7} \\
\times 13^{-6}17^{-1}\\
\times  (20891, 18501);  \\
 \end{array} $  & 2.236532215249387E-004     \\  \hline

$\left\{ \begin{array}{llllllll} 5 && 5 && 4 && 3 & \\ & 6 && 4 && 6 && 4\\
2 && 4 && 7 && 5 & \end{array} \right\} $
 & $\begin{array}{l} 2^{-16}3^{-6}5^{3}
 7^{-7}11^{-5} \\
\times 13^{-5}19^{1}109^{1}521^{2}\\
\times  (2467);  \\
 \end{array} $  & -4.451211337906913E-003     \\  \hline
 $\left\{ \begin{array}{llllllll} 6 && 4 && 7 && 4 & \\ & 2 && 5 && 7 && 6\\
4 && 5 && 6 && 3 & \end{array} \right\} $
 & $\begin{array}{l} 2^{-1}3^{-8}5^{-3}
 7^{-6}11^{-6} \\
\times 13^{-7}17^{-2}19^{-1}\\
\times  (10963,3214,29879);  \\
 \end{array} $  & 3.429876771358886E-002     \\  \hline
 $\left\{ \begin{array}{llllllll} 7 && 8 && 9 && 10 & \\ & 8 && 6 && 4 && 6\\
7 && 9 && 7 && 5 & \end{array} \right\} $
 & $\begin{array}{l} -2^{-30}3^{-3}5^{3}7^{-3}11^{-9}\\
\times 13^{-5} 17^{-8}19^{-7}\\
\times 23^{-6}41^{2}1153^{2}\\
\times (3297, 6491, 22005,\\
\times 31646, 20367);  \\
 \end{array} $  & -0.708747948235219     \\  \hline
 $\left\{ \begin{array}{llllllll} 10 && 7 && 8 && 6 & \\ & 9 && 10 && 6 && 8\\
7 && 9 && 10 && 7 & \end{array} \right\} $
 & $\begin{array}{l} -2^{7}3^{-7}5^{-9}
 7^{-3}11^{-7}\\
\times 13^{-11} 17^{-8}19^{-8}29^{1}47^{2}\\
\times 179^{2}\times (42,6600,\\
\times 31323, 17916, 4465);  \\
 \end{array} $  & -0.320311506498916     \\  \hline
 $\left\{ \begin{array}{llllllll} 20 && 15 && 9 && 10 & \\ & 14 && 18 && 15 && 15\\
9 && 8 && 10 && 12 & \end{array} \right\} $
 & $\begin{array}{l} 2^{3}3^{-13}5^{-19}
 7^{-3} 11^{-5} \\
\times 17^{-6}19^{1}23^{-4}29^{-8}\\
\times 31^{-8} 37^{-1}41^{-1}43^{-1}\\
\times 263^{2}\times (20491, 6148,\\
\times 19764, 6941);  \\
 \end{array} $  & 1.315510402031053E-013     \\  \hline\hline
   \end{tabular}
\end{center}
\end{table}

\begin{table}[p]
\begin{center}
  \begin{tabular}{|l|l|}  \hline\hline
 $\left\{\begin{array}{cccccccc}j_{1}&&j_{6}&&j_{7}&&j_{8}&\\
&j_{2}&&j_{3}&&j_{4}&&j\\j_{5}&&j^{'}_{6}&&j^{'}_{7}&&j^{'}_{8}&\end{array}\right\}$
& $\left\{\begin{array}{cccccccc}j_{1}&&j_{6}&&j_{7}&&j_{8}&\\
&j_{2}&&j_{3}&&j_{4}&&j\\j_{5}&&j^{'}_{6}&&j^{'}_{7}&&j^{'}_{8}&\end{array}\right\}$
\\ \hline\hline
$\left\{ \begin{array}{llllllll} 3 && 2 && 2 && 2 & \\ & 2 && 1 && 0 && 3\\
2 && 1 && 0 && 0 & \end{array} \right\} $ & $
\left\{ \begin{array}{llllllll} 2 && 3 && 2 && 2 & \\ & 2 && 2 && 0 && 2\\
1 && 2 && 0 && 0 & \end{array} \right\} $
\\  \hline
$\left\{ \begin{array}{llllllll} 2 && 2 && 3 && 3 & \\ & 1 && 2 && 0 && 2\\
2 && 2 && 0 && 0 & \end{array} \right\} $ & $
\left\{ \begin{array}{llllllll} 2 && 2 && 1 && 1 & \\ & 3 && 2 && 0 && 2\\
2 && 2 && 0 && 0 & \end{array} \right\} $
\\  \hline
$\left\{ \begin{array}{llllllll} 2 && 1 && 2 && 2 & \\ & 2 && 2 && 0 && 2\\
3 && 2 && 0 && 0 & \end{array} \right\} $ & $
\left\{ \begin{array}{llllllll} 1 && 2 && 2 && 2 & \\ & 2 && 3 && 0 && 1\\
2 && 3 && 0 && 0 & \end{array} \right\} $
\\  \hline
$\left\{ \begin{array}{llllllll} 3 && 2 && 2 && 2 & \\ & 2 && 0 && 1 && 3\\
2 && 1 && 1 && 0 & \end{array} \right\} $ & $
\left\{ \begin{array}{llllllll} 3 && 2 && 1 && 2 & \\ & 2 && 1 && 1 && 3\\
2 && 1 && 1 && 0 & \end{array} \right\} $
\\  \hline
$\left\{ \begin{array}{llllllll} 3 && 2 && 2 && 2 & \\ & 2 && 1 && 1 && 3\\
2 && 1 && 1 && 0 & \end{array} \right\} $ & $
\left\{ \begin{array}{llllllll} 3 && 2 && 3 && 2 & \\ & 2 && 1 && 1 && 3\\
2 && 1 && 1 && 0 & \end{array} \right\} $
\\  \hline
$\left\{ \begin{array}{llllllll} 3 && 2 && 1 && 2 & \\ & 2 && 2 && 1 && 3\\
2 && 1 && 1 && 0 & \end{array} \right\} $ & $
\left\{ \begin{array}{llllllll} 3 && 2 && 2 && 2 & \\ & 2 && 2 && 1 && 3\\
2 && 1 && 1 && 0 & \end{array} \right\} $
\\  \hline
$\left\{ \begin{array}{llllllll} 3 && 2 && 3 && 2 & \\ & 2 && 2 && 1 && 3\\
2 && 1 && 1 && 0 & \end{array} \right\} $ & $
\left\{ \begin{array}{llllllll} 2 && 3 && 2 && 2 & \\ & 2 && 2 && 1 && 2\\
0 && 2 && 1 && 0 & \end{array} \right\} $
\\  \hline
$\left\{ \begin{array}{llllllll} 1 && 3 && 2 && 2 & \\ & 2 && 2 && 1 && 1\\
1 && 2 && 1 && 0 & \end{array} \right\} $ & $
\left\{ \begin{array}{llllllll} 1 && 3 && 2 && 2 & \\ & 3 && 2 && 1 && 1\\
1 && 2 && 1 && 0 & \end{array} \right\} $
\\  \hline
$\left\{ \begin{array}{llllllll} 2 && 3 && 2 && 2 & \\ & 2 && 1 && 1 && 2\\
1 && 2 && 1 && 0 & \end{array} \right\} $ & $
\left\{ \begin{array}{llllllll} 2 && 3 && 3 && 2 & \\ & 2 && 1 && 1 && 2\\
1 && 2 && 1 && 0 & \end{array} \right\} $
\\  \hline
$\left\{ \begin{array}{llllllll} 2 && 3 && 2 && 2 & \\ & 1 && 2 && 1 && 2\\
1 && 2 && 1 && 0 & \end{array} \right\} $ & $
\left\{ \begin{array}{llllllll} 2 && 3 && 1 && 2 & \\ & 2 && 2 && 1 && 2\\
1 && 2 && 1 && 0 & \end{array} \right\} $
\\  \hline\hline

\end{tabular}
\end{center}
\end{table}

\end{document}